\documentstyle[times,epsfig]{l-aa}
\newcommand{\vavecA}{\bar{\vec{v}}_{\rm A}}
\newcommand{\va}{v_{\rm A}}
\newcommand{\vaA}{\bar{v}_{\rm A}}
\newcommand{\kvec}{\vec{k}}
\newcommand{\xvec}{\vec{x}}
\newcommand{\cscs}{c_{\rm s}^2}
\newcommand{\cs}{c_{\rm s}}
\newcommand{\Bvec}{\vec{B}}

\newcommand{\vvec}{\vec{v}}
\newcommand{\vvecA}{\bar{\vvec}}

\newcommand{\dt}[1]{\frac{\partial #1}{\partial t}}
\newcommand{\drho}{\delta\rho}
\newcommand{\dvvec}{\delta\vvec}
\newcommand{\dBvec}{\delta\Bvec}
\newcommand{\BvecA}{\bar{\Bvec}}
\newcommand{\rhoA}{\bar{\rho}}
\newcommand{\Fradvec}{\vec{F}_{\rm rad}}
\newcommand{\er}{\vec{e}_r}
\newcommand{\ep}{\vec{e}_\phi}
\newcommand{\Frad}{F_{\rm rad}}
\newcommand{\Mdot}{\dot{M}}
\newcommand{\kcak}{k_{\rm CAK}}
\newcommand{\acak}{\alpha_{\rm CAK}}
\newcommand{\Lsobo}{L_{\rm Sobo}}
\newcommand{\Msun}{M_\odot}
\newcommand{\Rsun}{R_\odot}
\newcommand{\vp}{v_{\rm Ph}}
\newcommand{\complexi}{{\rm i}}
\newcommand{\vinf}{v_\infty}
\newcommand{\vth}{v_{\rm th}}
\newcommand{\vrA}{\bar{v}_r}
\newcommand{\RA}{R_{\rm A}}
\newcommand{\vinfobs}{v_{\infty\rm,obs}}

\newcommand{\tauM}{\tau_M}
\newcommand{\tauJ}{\tau_J}
\newcommand{\chiA}{\bar{\chi}}
\newcommand{\yr}{\rm yr}
\begin{document}
\thesaurus{07(08.01.3; 08.13.1; 08.13.2; 08.18.1; 08.05.1)}
\title{Unstable waves in winds of magnetic massive stars}
\author{Henning Seemann \and Peter L.\ Biermann}
\institute{Max-Planck-Institut f\"ur Radioastronomie, Auf dem H\"ugel 69,
53121 Bonn, Germany}
\offprints{seemann@mpifr-bonn.mpg.de}
\date{Received November 25, 1996; accepted June 2, 1997}
\maketitle
\begin{abstract}
We use a luminous fast magnetic rotator model to analyze the influence of a
magnetic field on the linear waves induced in the wind of a massive star by the
radiative instability. We show that a twisted magnetic field can drive a strong
wind with a wind efficiency $(\Mdot\vinf)/(L/c)>1$ even without multiple
scattering. The radiation amplified waves in the wind are modified by the
twisted magnetic field so that they can enhance the wind and lead to
overestimates for $\Mdot$ and $\vinf$. Finally we argue that the spin down time
might be consistent with the lifetime derived from the mass loss rate within
the uncertainties regarding the stellar structure. Therefore our model may help
to explain high, observed values for $\Mdot$ and $\vinf$ without being ruled
out by the spin down problem.  
\end{abstract}
\section{Introduction}
The wind of massive stars is primarily driven by the interaction with stellar
radiation in many lines (Lucy \& Solomon \cite{r:luc:sol}). Based on this
concept, the theory of Castor, Abbott, and Klein (\cite{r:cas:abb:kle}
hereafter CAK) can explain many aspects of hot star winds.  But the basic model
of CAK fails to describe the strong winds observed in some massive stars,
including Wolf-Rayet stars.  Beside multiple scattering effects (Gayley et al.\
\cite{r:gay:owo:cra}) a rotating magnetic field as driving force were proposed
(Friend \& MacGregor \cite{r:fri:mcg}, Poe et al.\ \cite{r:poe:fri:csi},
Maheswaran \& Cassinelli \cite{r:mar:csi:92}, and Biermann \& Cassinelli
\cite{r:bie:csi}) to overcome this problem. 

 We emphasize, that the observations of nonthermal radio emission from OB
(Bieging et al.\ \cite{r:big:abb:chu}) and Wolf-Rayet stars (Abbott et al.\
\cite{r:abb:big:chu:tor}) give clear evidence for non neglible magnetic fields
on these stars. On the other hand direct observations give only rather high
limits on the magnetic field (Landstreet \cite{r:lnd}), especially for very
early-type stars. Landstreet's sample contains only one O-star (O9.5), which is
not massive enough to form a Wolf-Rayet star later. Further indirect hints for
a strong magnetic field come from radio observations of supernova explosions
(Biermann \& Cassinelli \cite{r:bie:csi}, Biermann et al.\
\cite{r:bie:str:fal}, Biermann \cite{r:bie:97}). More observations (e.g.\
Ignace et al.\ \cite{r:ign:nor:csi}) are necessary, before our model can be
verified or ruled out.

Previous models for rotating magnetic stars suffer from one serious problem: A
sufficient enhancement of the wind is connected to a severe loss of angular
momentum. The star would spin down faster than is allowed from stellar
evolution. But Maheswaran \& Cassinelli (\cite{r:mar:csi:94}) argued that a
massive star could even reach the Wolf-Rayet phase as a fast magnetic
rotator. Magnetic fields may also play an important role in rotationally
compressed winds (Cassinelli et al.\ \cite{r:csi:ign:bjo}).

Another important feature of hot star winds is the instability of the
radiation force, which leads to a highly perturbed wind.  The stability and
propagation of linear hydrodynamical waves in the radiation field of the star
were analyzed by Owocki \& Rybicki (\cite{r:owo:ryb:84} hereafter OR, and
\cite{r:owo:ryb:85}), who unified prior work by MacGregor et al.\
(\cite{r:mcg:har:ray}) and Abbott (\cite{r:abb}). They found that the linear
waves grow very rapidly and therefore cannot be neglected in an overall
model. The resulting waves can influence the stellar wind in two ways: (i)
They can gain momentum and energy from the radiation field, transport it, and
later dissipate it into the wind. This can help to drive the unperturbed wind
(Koninx \cite{r:kon}) e.g.\ by increasing the real $\Mdot$ at the base of the
wind, where shocks might not have formed yet. (ii) Waves can also steepen into
shocks and influence the observation of fundamental wind parameters: The {\it
observed} terminal velocity $\vinf$ is inferred from the blue edge of P-Cygni
lines. This {\it maximal} velocity has to be compared with the unperturbed
$\vinf$ plus the velocity amplitude of the waves or, if {\it outward} running
waves have already steepened into outward running shocks, to the unperturbed
$\vinf$ plus the shock speed. The last scenario is more probable due to the
high amplification rates found in OR\@. Krolik \& Raymond
(\cite{r:kro:ray}) and MacFarlane \& Cassinelli (\cite{r:mcf:csi}) analyzed
the structure of a single outward running shock shell in a nonmagnetic wind
and found that the shock shell is driven by the radiation field to a velocity
much higher than the velocity of sound. The {\it observed} values for the mass
loss rate $\Mdot$ are influenced by waves in two ways: (i) $\Mdot$ is inferred
from observed values of the density $\rho$ by $\Mdot \sim \rho r^2 \vinf$. An
overestimated $\vinf$ therefore leads to an overestimated $\Mdot$. (ii) The
values for $\rho$ are inferred from observed radio or UV fluxes. A matter
distribution perturbed by waves or shocks generates higher fluxes, which lead
to higher values for $\rho$ if the perturbations are not taken into account
properly (Abbott et al.\ \cite{r:abb:big:chu} and Hillier
\cite{r:hil}). Shocks are also a good source for the observed X-rays from hot
stars.  Lucy (\cite{r:luc:82}) developed a heuristic model consisting of a
chain of outward running forward shocks mainly in order to describe the X-ray
emission from hot stars.  But Owocki et al.\ (\cite{r:owo:cas:ryb}) found in a
nonlinear time-dependent calculation, that waves in the stellar wind are
dominantly running inward and steepen into inward running, reverse shocks. In
this case the arguments about $\vinf$ and also Koninx's model do not apply, so
that such waves can not significantly help to explain the high values for
$\Mdot$ and $\vinf$ observed in the wind of massive stars.

The aim of this paper is to show in the limit of a linear analysis that the
objections of Owocki et al.\ (\cite{r:owo:cas:ryb}) do not apply, if a strong
magnetic field and rotation are present. Therefore we join the ideas of a
rotating magnetic field (Maheswaran \& Cassinelli
\cite{r:mar:csi:92},\cite{r:mar:csi:94}) and of unstable waves (Owocki \&
Rybicki \cite{r:owo:ryb:84}(OR)) in the wind of a massive star and do a linear
stability analysis. We show that the magnetic field increases the number of
wave modes and changes their properties significantly. High phase velocities
can be achieved due to the high Alfv\'en velocity. Therefore inward running
waves will not be advected outward and will not steepen into reverse shocks
anymore, which dominate at large radii.  Furthermore outward and inward running
waves have the same growth timescale in the short wavelength regime where both
modes are unstable. So we can expect outward running waves far from the star
and forward shocks in the nonlinear regime. This model does not exclude
multiple scattering as described e.g.\ by Lucy \& Abbott (\cite{r:luc:abb}) or
Gayley et al.\ (\cite{r:gay:owo:cra}). Rather both could be joined to create a
model for an even stronger wind.

In Sect.~2 we derive the dispersion relation for radiatively amplified waves in
the presence of a magnetic field. In Sect.~3 we discuss our unperturbed wind
models. In Sect.~4 we discuss the waves we found for the wind models of
Sect.~3. In Sect.~5 we discuss the observational consequences of our model. And
in Sect.~6 we describe some conclusions.
\section{The wave equations}
To analyze waves in the wind of hot stars we start from the equations of
magnetohydrodynamics for a compressible, nonviscous, perfectly conducting fluid
as described in Jackson (\cite{r:jac}) and add the analytic description of
OR for the influence of the stellar radiation on the plasma. OR was
criticized by Lucy (\cite{r:luc:84}) for the neglect of damping due to diffuse
radiation. But Owocki \& Rybicki (\cite{r:owo:ryb:85}) showed that this effect
reduces the amplification rate only by approximately 50\% at one stellar radius
and 20\% at infinity. In this initial paper we emphasize the effect of the
magnetic field. Therefore we choose the simple description of OR instead
of the more exact but also more involved description of Owocki \& Rybicki
(\cite{r:owo:ryb:85}) or the description of Gayley \& Owocki
(\cite{r:gay:owo}), who analyzed the instability in optically thick winds. We
start with
\begin{eqnarray}\label{basiceqn1}
0 &=& \dt{\rho} + \nabla\cdot(\rho\vvec)\\ 
0 &=& \rho\dt{\vvec} + \rho(\vvec\cdot\nabla)\vvec + \nabla p + 
\frac{\Bvec}{4\pi}\times(\nabla\times\Bvec) - \Fradvec\\
0 &=& \dt{\Bvec} - \nabla\times(\vvec\times\Bvec)\label{basiceqn3}
\end{eqnarray}
where $\Fradvec$ is the force term due to radiation pressure. Now we express
the pressure $p$ by $\cscs\rho$, where $\cs$ is the speed of sound, and
replace $\rho$, $\vvec$, and $\Bvec$ by sums of a equilibrium value and a
perturbation. In the comoving reference frame $(\vvecA=0)$\footnote{Variables
marked with a\ $\bar{\ }$\ refer to the unperturbed wind.} this is:
\begin{eqnarray}
\rho  & = & \rhoA  + \drho(\xvec,t) \\
\vvec & = & \dvvec(\xvec,t) \\ 
\Bvec & = & \BvecA + \dBvec(\xvec,t)
\end{eqnarray}
Equations \ref{basiceqn1}--\ref{basiceqn3} give to first order in small
quantities:
\begin{eqnarray}\label{lineqn1}
0 &=& \dt{\drho(\xvec,t)} + \rhoA\nabla\cdot\dvvec(\xvec,t) \\
0 &=& \rhoA\dt{\dvvec(\xvec,t)} + \cscs\nabla\drho(\xvec,t) + 
\frac{\BvecA}{4\pi}\times(\nabla\times\dBvec(\xvec,t))\nonumber\\
&& -{}\delta\Fradvec\\
0 &=& \dt{\dBvec(\xvec,t)} - \nabla\times(\dvvec(\xvec,t)\times\BvecA)
\label{lineqn3}
\end{eqnarray}
We assume now, that the radiative force acts only in radial direction and
depends only on the radial velocity.  These equations can be reduced to a
single equation for $\dvvec$:
\begin{eqnarray}
0 &=& \frac{\partial^2\dvvec(\xvec,t)}{\partial t^2} - \cscs\nabla(\nabla\cdot
\dvvec(\xvec,t)) +\nonumber\\
&&\vavecA\times\nabla\times[\nabla\times(\dvvec(\xvec,t)\times\vavecA)]
\nonumber\\
&&-{}\frac{1}{\rhoA}\frac{\partial\delta v_r(\xvec,t)}{\partial t} 
\frac{\partial \Frad}{\partial v_r}\er
\label{waveeqn}
\end{eqnarray}
with the vectorial Alfv\'en velocity $\vavecA = \BvecA/\sqrt{4\pi\rhoA}$. If we
use the result of OR for the linear perturbation of the radiative force,
we can get a dispersion relation from this wave equation for plane waves:
\begin{equation}
\dvvec(\xvec,t) = \dvvec e^{\complexi\kvec\cdot\xvec - \complexi \omega t}.
\end{equation}
Equation \ref{waveeqn} then becomes:
\begin{eqnarray}\label{vecdisp}
0&=&-\omega^2\dvvec + (\cscs + \va^2)(\kvec\cdot\dvvec)\kvec + 
(\vavecA\cdot\kvec)[(\vavecA\cdot\kvec)\dvvec\nonumber\\
&&-{}(\vavecA\cdot\dvvec)\kvec - (\kvec\cdot\dvvec)\vavecA]
-\frac{\Omega k_r}{\chiA + \complexi k_r}\omega (\dvvec\cdot\er)\er
\end{eqnarray}
Here we used the analytic form of the perturbation of the radiation force
derived in OR.
\begin{equation}
\Omega=\sqrt{\frac{2c^{\acak-1/2}}{1-\acak}}\frac{\vrA}{\Lsobo}
\end{equation}
is their combined amplification rate of all lines. 
\begin{equation}
\chiA=\sqrt{\frac{2c^{\acak-1/2}}{1-\acak}}\frac{1}{\Lsobo}=\frac{\Omega}{\vrA}
\end{equation}
is their mean blue-edge absorption strength with the empirical parameter
$c\approx 1.6$. And $\vrA$ is the radial velocity of the unperturbed wind in
the rest frame of the star.  $\Lsobo=\vth/(\partial \vrA/\partial r)$ is the
Sobolev length. Equation
\ref{vecdisp} is a vector equation, which is linear in $\dvvec$. We can think
of it as a generalized eigenvalue problem:
\begin{equation}\label{eigeneqn}
(\tens{A}(\kvec,\vavecA,\cscs) - \omega \tens{B}(k_r,\Omega,\chiA) -
\omega^2\bbbone)\dvvec =0
\end{equation}
We can find $\omega$ and $\dvvec$ numerically.  Then $\dBvec$ and $\drho$
follow from Eqs. \ref{lineqn1} \& \ref{lineqn3}:
\begin{eqnarray}
\drho  & = & \frac{\rhoA}{\omega}(\kvec\cdot\dvvec)\\
\dBvec & = & \frac{1}{\omega}[(\kvec\cdot\dvvec)\BvecA - 
(\BvecA\cdot\kvec)\dvvec\,]
\end{eqnarray}
Although Eq.~\ref{eigeneqn} is very involved in the general case, we can
find an analytical solution for a simplified situation:
\begin{eqnarray}
\kvec(r) &=& k \er\\
\vavecA(r) &=& \vaA \ep
\end{eqnarray}
The latter approximation is quite accurate far away from the star. In this
limit we find
\begin{equation}
\omega=-\frac{\Omega k}{2(\chiA+\complexi k)}\pm\sqrt{\left(
\frac{\Omega k}{2(\chiA+\complexi k)}\right)^2 + (\cscs+\vaA^2)k^2}.
\end{equation}
In the long wavelength limit $(k \ll \chiA)$, where the waves are stable, this
leads to
\begin{equation}
\omega=\left[-\frac{\vrA}{2}\pm\sqrt{\frac{\vrA^2}{2}+\cscs+\vaA^2}\,\right] k.
\end{equation}
In the case of a weak magnetic field with $\vrA \gg \cs \ga \vaA$, this
resembles Abbott's (\cite{r:abb}) result for stable radiative-acoustic waves
with a fast inward and a slow outward mode. In the case of a strong magnetic
field we have $\vrA\approx\vaA\gg\cs$.  This leads to higher phase velocities
for both modes and reduces the relative difference between inward and outward
waves. Additionally inward running waves are not advected outward by the
average wind motion any more, because their phase velocity is higher than the
velocity of the unperturbed wind. In the short wave length limit $(k \gg
\chiA)$ we find
\begin{equation}
\omega=\complexi\frac{\Omega}{2}\pm\sqrt{-\frac{\Omega^2}{4}+
(\cscs+\vaA^2)k^2}.
\end{equation}
These waves propagate, if $\Omega$ is less than $2k\sqrt{\cscs+\vaA^2}$, with
the same phase velocity inward and outward. The amplification rate is also
the same for both modes. 

In the case of no magnetic field Eq.~\ref{eigeneqn} leads to
\begin{equation}\label{dispnB}
0=\omega^3+\frac{\Omega k_r}{\chiA + \complexi k_r}\omega^2-\cscs\kvec^2
\omega-\frac{\Omega k_r}{\chiA + \complexi k_r}\cscs(k_\phi^2 + k_\theta^2).
\end{equation}
For $k_\theta=k_\phi=0$ this reproduces the result for isothermal waves found
in OR. For oblique waves there is a third wave mode.
\section{The unperturbed wind models}
To analyze the effect of these waves we construct three wind models for a
standard massive star. Table \ref{t:svalues} gives the fundamental
parameters of our standard star. 
\begin{table}
\caption[]{\label{t:svalues}Fundamental parameters of our standard
star}
\begin{flushleft}
\begin{tabular}{ll}
\noalign{\smallskip}
\hline 
\noalign{\smallskip}
$M$ & $23M_\odot$\\ 
$L$ & $1.7 \times 10^5L_\odot$\\ 
$R$ & $8.5\Rsun$\\ 
$T$ & $60000$K\\ 
$\acak$ & 0.56\\ 
$\kcak$ & 0.28\\
\noalign{\smallskip}
\hline 
\end{tabular}
\end{flushleft}
\end{table}
The first model (model~A) is the standard analytic CAK wind for a star without
magnetic field, rotation, or pressure. In this model we reproduce the previous
result of dominantly inward running waves found by Owocki et
al.~(\cite{r:owo:cas:ryb}). In the second model (model~B) we add a radial
magnetic field. Since this magnetic field is parallel to the natural stream
lines of the wind of a nonrotating star, this field does not change the
velocity profile of the unperturbed wind. But it changes the microphysics for
waves. The last model (model~C) is a luminous fast magnetic rotator model.  The
rotating magnetic field provides an additional driving force, which changes the
properties of the wind drastically. The terminal velocity decreases and the
mass loss rate increases. The wind efficiency $(\Mdot\vinf)/(L/c)$ is 3.8,
which is much higher than for a purely radiatively driven wind in the single
scattering limit, where the efficiency can not be higher than unity. In spite
of the known spin down problem we choose a strong magnetic field and a high
rotation rate in order to emphasize the influence of these parameters. For this
model we use the equations derived by Biermann \& Cassinelli
(\cite{r:bie:csi}), which do not assume that the magnetic field is radial close
to the star. This leads to a smaller and more accurate value for the Alfv\'enic
radius $\RA$, which controls the angular momentum loss of the star.  Table
\ref{t:mvalues} gives the magnetic field, rotation rate, and the resulting
values for the above mentioned unperturbed wind models. In Sect.~5 we will
discuss how these values and their observation can be influenced by waves. In
this paper we do not discuss the influence of the waves back on the unperturbed
wind as Koninx (\cite{r:kon}) did. A model with rotation but without magnetic
field is not included, because this model has the same microphysics for waves
as model~A\@. Just $\Mdot$ and the velocity dependence on $r$ are
different. Model~C differs from model~B in the local conditions by the fact
that, due to the rotational twist, the magnetic field and the direction of the
radiative force are not parallel anymore. Even at the base of the wind $B_\phi$
is approximately $-1.67B_r$.
\begin{table}
\caption[]{\label{t:mvalues}Results for the unperturbed wind models}
\begin{flushleft}
\begin{tabular}{lrrrrr}
\noalign{\smallskip}
\hline 
\noalign{\smallskip}
Model                                  &    A &    B &    C \\
\noalign{\smallskip}
\hline 
\noalign{\smallskip}
$B_{r0}\ [{\rm G}]$                    &    0 &  500 &  500 \\
$\Omega/\Omega_{\rm crit}$             &    0 &    0 & 0.92 \\
$\vinf\ [{\rm km\ s^{-1}}]$            & 1146 & 1146 &  784 \\
$\Mdot\ [10^{-6}\Msun\ {\rm yr}^{-1}]$ &  0.6 &  0.6 &   17 \\
$R_{\tau=2/3}\ [\Rsun]$                &  8.5 &  8.5 &   11 \\
$(\Mdot\vinf)/(L/c)$                   &  0.2 &  0.2 &  3.8 \\
\noalign{\smallskip}
\hline 
\end{tabular}
\end{flushleft}
\end{table}
\section{Numerical results for waves}
Since our unperturbed wind model is limited to the equatorial plane we limit
our discussion for waves to the same plane. We have still two free parameters
then: The wavelength and the azimuthal angle of $\kvec$. In the first section
we discuss radial waves. This will show all important properties of this wave
model. In the second section we discuss briefly the influence of the azimuth
angle of $\kvec$.
\subsection{Radial Waves}
\begin{figure*}
\setlength{\unitlength}{1cm}
\begin{picture}(18,8.7)
\put(0,0){\epsfig{figure=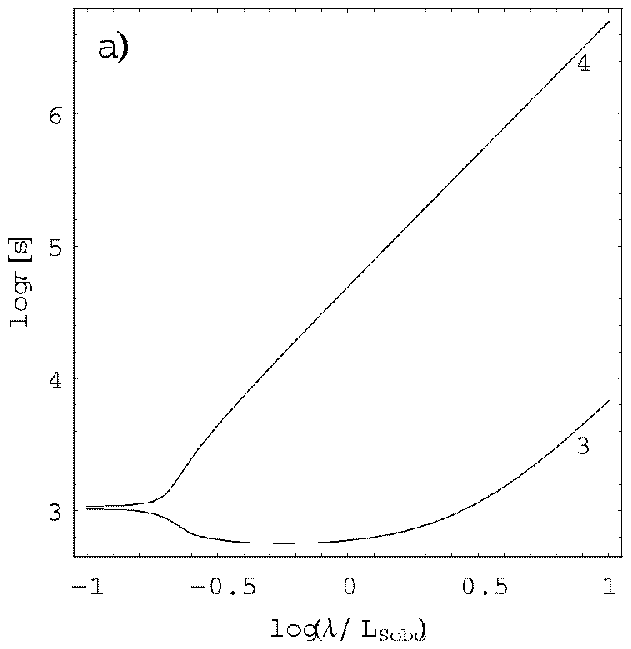}}
\put(9,0){\epsfig{figure=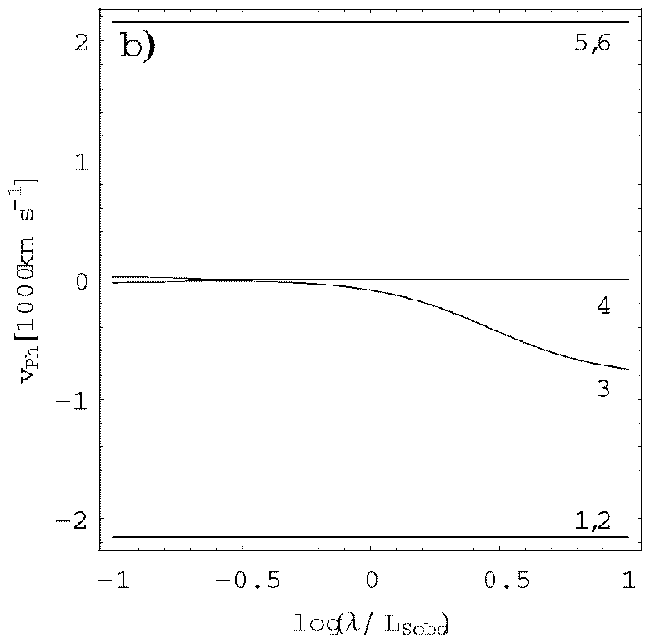}}
\end{picture}
\caption[]{\label{f:moda:a}Amplification timescales (a) and phase velocities 
(in the frame of the unperturbed wind) (b) versus
wavelength for model~A\&B and $r=2R$. In model~A only modes~3\&4 exist. Modes 
missing in Fig.~a) are stable. Mode~3 with $\lambda\approx\Lsobo$ has the 
shortest amplification timescale and therefore will dominate the wind. These 
inward running waves are advected outward with $\vrA=811\rm km\ s^{-1}$ and 
steepen into reverse shocks.}
\end{figure*}
\begin{figure*}
\setlength{\unitlength}{1cm}
\begin{picture}(18,8.7)
\put(0,0){\epsfig{figure=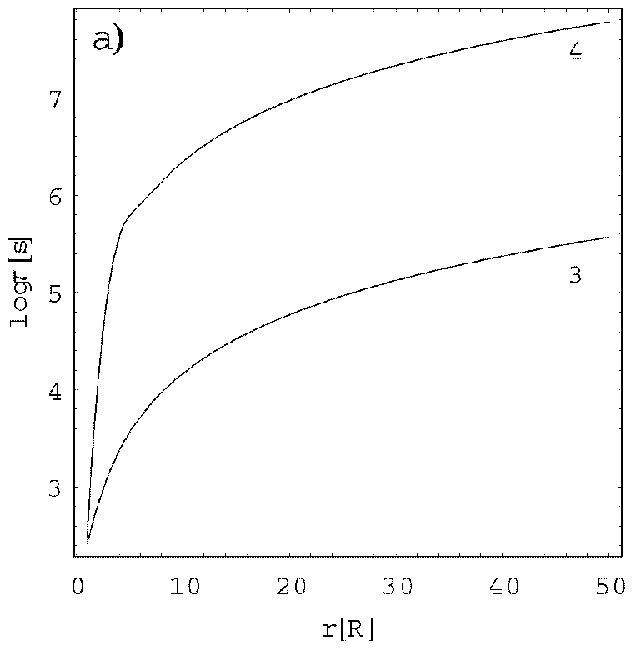}}
\put(9,0){\epsfig{figure=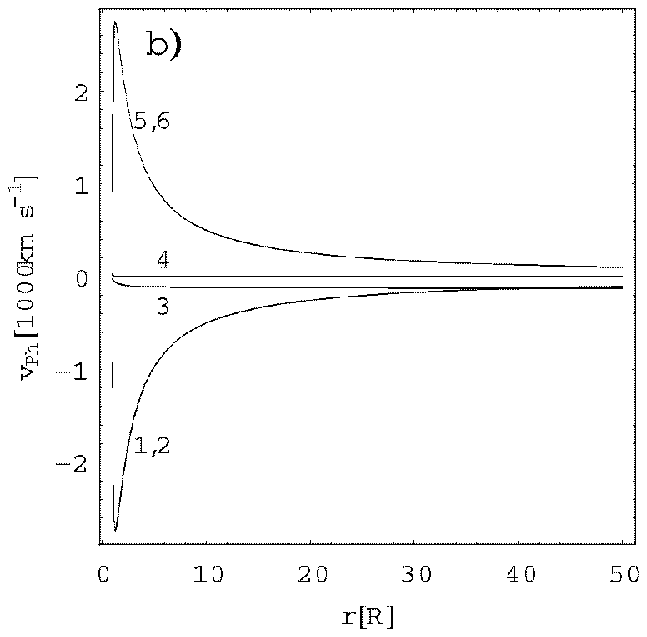}}
\end{picture}
\caption[]{\label{f:moda:b}Amplification timescales (a) and phase velocities 
(in the frame of the unperturbed wind) (b) versus radius 
for model~A\&B and $\lambda=\Lsobo$. In model~A only modes~3\&4 exist. 
Modes missing in Fig.~a) are stable. The wave amplification is strongest close
to the star, where the radiation field and wind acceleration are strong. Inward
running waves originating there will be advected outward and steepen into
reverse shocks.}
\end{figure*}
Figures \ref{f:moda:a}\&\ref{f:moda:b} show numerical results for wind model
A\&B. The sonic wave modes\footnote{Ordered by the radial phase velocity we
denote in our figures the fast magnetosonic modes with 1\&6, the Alfv\'enic
modes with 2\&5, and the slow magnetosonic modes with 3\&4. The sonic modes of
model~A, which are identical to the slow magnetosonic modes of model~B, are
denoted with 3\&4 as well.} found for model~A are identical to the slow
magnetosonic modes of model~B\@. For model~B we find additionally four fast
wave modes. These modes, the fast magnetosonic and the Alfv\'enic modes, have
the same phase velocity $\vp \approx \vaA$, because both, $\Bvec$ and $\kvec$,
are parallel to $\er$.  But these wave modes are stable and show no dependence
on wavelength. Therefore we concentrate our discussion for model A\&B on the
slow waves: Fig.~\ref{f:moda:a} shows the dependence of the waves on the
wavelength for $r=2R$, plotted relative to the Sobolev length, which is the
relevant length scale for radiative wave amplification in the model of OR,
we use here. $R=8.5\Rsun$ is the stellar radius. In the long wavelength limit
we find stable waves with a high phase velocity inward and a low phase velocity
outward. This reproduces Abbott's (\cite{r:abb}) result of radiative-acoustic
waves. In the short wavelength limit we find the same amplification timescales
and phase velocities (except the direction) for both modes. In this limit we
would expect to see both wave modes in the wind. But the most interesting case
is the bridging case $\lambda \approx\Lsobo$. Here we find the shortest
amplification timescale for all wavelengths. Inward waves are two orders of
magnitude faster in amplification than outward waves. They have also a higher
phase velocity. This resembles the result of Owocki et al.\
(\cite{r:owo:cas:ryb}), who did a nonlinear calculation and found that
primarily the inward running waves steepen into reverse shocks, which are
advected outward. The aim of this paper is to argue that this scenario changes
if a magnetic field {\em and} rotation are involved. Figure \ref{f:moda:b}
shows the radial dependence of the wave with $\lambda = \Lsobo(r)$. The phase
velocity for the slow magnetosonic modes is approximately the velocity of
sound, so that inward running waves are advected outward in the supersonic part
of the wind. The amplification of the waves is strongest close to the star,
where the Sobolev length is short. The phase velocity of the fast modes goes
with $r^{-1}$ for large radii since $B = B_r \sim r^{-2}$. This might lead to a
small velocity for outward running shocks at large radii.

\begin{figure*}
\setlength{\unitlength}{1cm}
\begin{picture}(18,8.7)
\put(0,0){\epsfig{figure=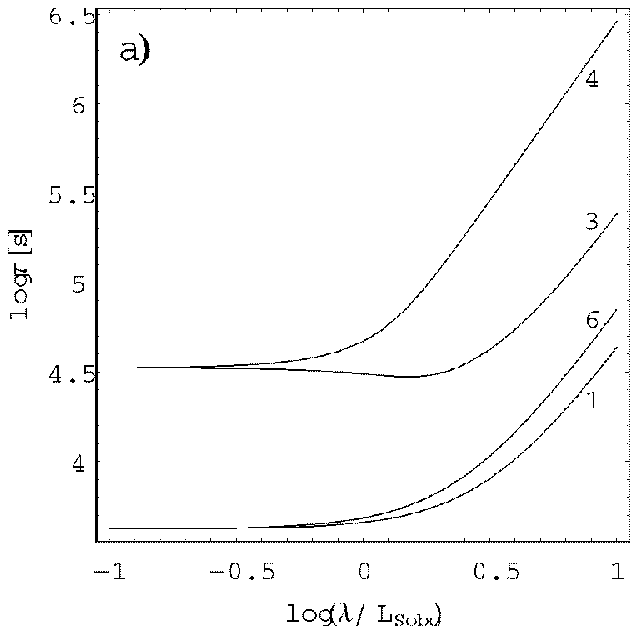}}
\put(9,0){\epsfig{figure=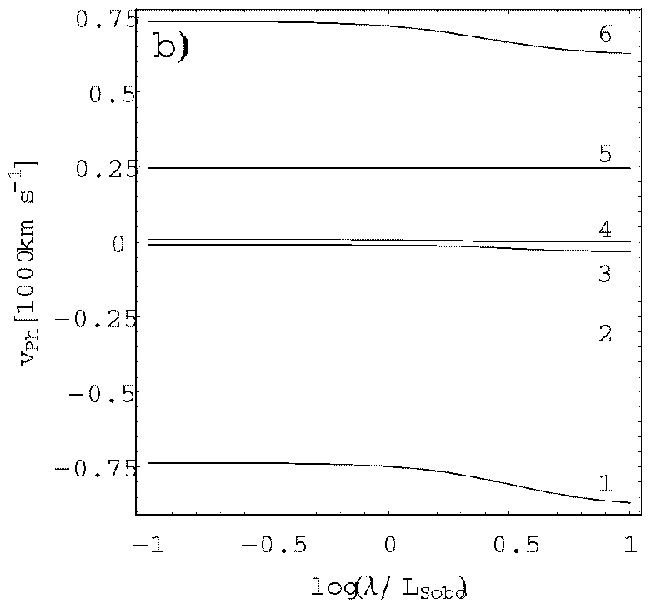}}
\end{picture}
\caption[]{\label{f:modb:a}Amplification timescales (a) and phase velocities 
(in the frame of the unperturbed wind) (b) versus wavelength for model~C and
$r=2R$. Modes missing in Fig.~a) are stable.  In the long wavelength limit all
modes are stable. In the short wavelength wavelength limit the fast
magnetosonic modes (1\&6) grow approximately one order of magnitude faster as
the slow magnetosonic modes (3\&4). We expect that fast magnetosonic waves in
both directions are equally dominant in the wind of model~C\@. Since the
magnetosonic waves are much faster than the unperturbed wind ($\vrA=294\rm km\
s^{-1}$), the inward running waves will not be advected outward. The stable
Alfv\'enic modes (2\&5) show no dependence on wavelength.}
\end{figure*}
\begin{figure*}
\setlength{\unitlength}{1cm}
\begin{picture}(18,8.7)
\put(0,0){\epsfig{figure=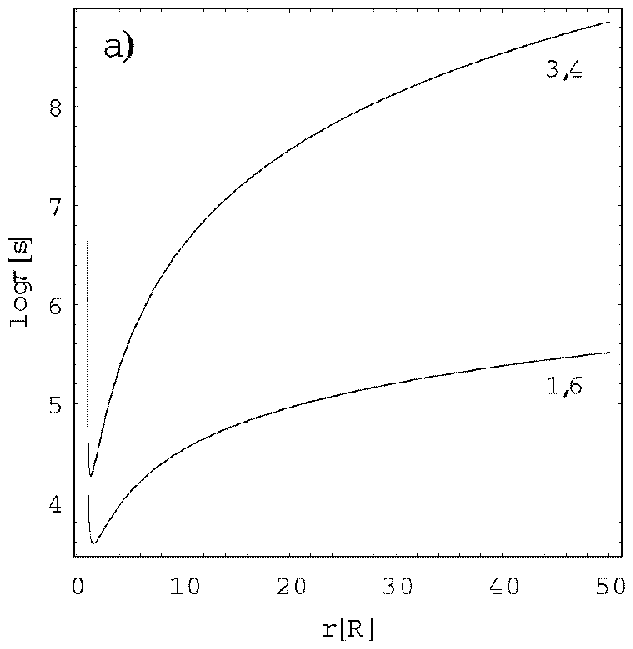}}
\put(9,0){\epsfig{figure=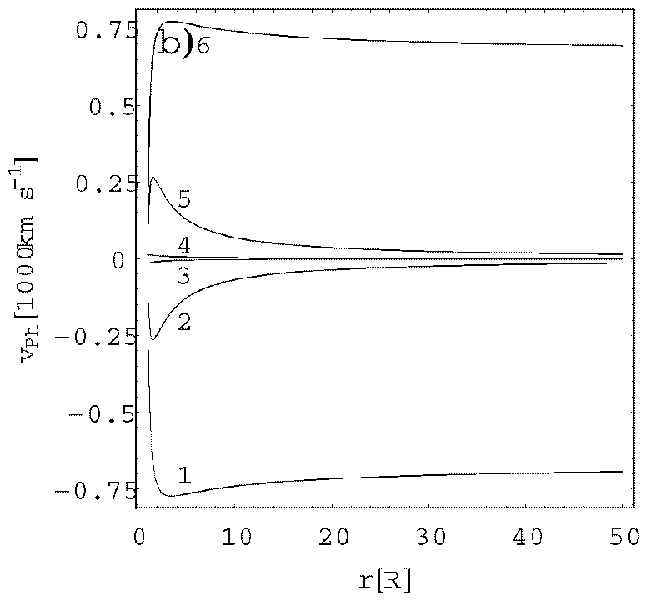}}
\end{picture}
\caption[]{\label{f:modb:b}Amplification timescales (a) and phase velocities 
(in the frame of the unperturbed wind) (b) versus radius for model~C and
$\lambda=0.1\Lsobo$. The Alfv\'enic modes (2\&5) missing in Fig.~a) are
stable. The wave amplification is strongest close to the star, where radiation
field and wind acceleration are strong. Fast magnetosonic waves originating
here will dominate the wind. They have the shortest amplification timescale for
all radii and a high phase velocity even at large radii. This may lead to fast
shocks running outward, which have a strong influence on observation.}
\end{figure*}
Figures \ref{f:modb:a}\&\ref{f:modb:b} show the same plots for model~C -- with
magnetic field and rotation. The crucial point is that the magnetic field and
the amplifying stellar radiation are not parallel anymore. For large radii they
are even perpendicular. Therefore the fast magnetosonic wave modes, which are
most interesting for us, are amplified as well. Figure \ref{f:modb:a}.b shows
six modes with different phase velocities. Two of them, the Alfv\'enic modes,
show no dependence on wavelength. They are unaffected by the radiation field
and therefore stable. Figure \ref{f:modb:a}.a shows the amplification
timescales for the magnetosonic waves. The fast magnetosonic waves grow
approximately one order of magnitude faster than the slow waves. We can
therefore expect that these waves will dominate. The crucial point is that they
are much faster than the unperturbed wind -- especially close to the
star. Inward running fast waves will therefore not be advected away from the
star. Figure \ref{f:modb:b}.b shows that the phase velocity for the fast
magnetosonic waves remain high for large radii. This velocity is a lower limit
for the velocity of outward running shocks. Rybicki et al.\
(\cite{r:ryb:owo:cas}) showed that nonradial perturbations in the stellar wind
are damped close to the wind. But the magnetic field of our model~C is mostly
tangential already at the stellar surface. For the fast magnetosonic modes
$|\delta v_\phi/\delta v_r|$ is 0.5 at the stellar surface and 0.35 at $r=2R$.
We expect therefore, that the effects found by Rybicki et al.\ will influence
but not completely dampen the magnetosonic waves. We emphasize that fast
magnetosonic waves propagate fastest perpendicular to the magnetic field with
$\vp=(\vavecA^2+\cscs)^{0.5}$.

It is very speculative to draw conclusions for nonlinear waves and shocks from
a linear stability analysis. But we showed that a magnetic field has a strong
influence on the linear analysis. Therefore we expect a strong influence of the
magnetic field on nonlinear perturbations as well.  Our speculative scenario
for waves in the wind of hot stars is the following: Waves are predominantly
generated close to the star where the stellar radiation field is strong and the
Sobolev length is short. Waves with a wavelength short compared to the Sobolev
length are generated predominantly, because they have the shortest
amplification timescale. Inward running waves run into the star and disappear
because their phase velocity is higher than the unperturbed wind
velocity. Outward running waves, which have, for short wavelengths, the same
growth timescale than inward running waves, can run over many stellar radii and
grow. For the outward running waves $\drho$ and $\delta v_r$ are in phase.
Therefore they can steepen into forward shocks with regions of high density at
high velocity and influence the measurements for the terminal velocity and mass
loss rate.
\subsection{Nonradial Waves}
The picture drawn in the last section does not change significantly when
nonradial waves are taken into account. There are two new effects compared to
the radial case. A third wave mode appears in model~A as predicted by
Eq.~\ref{dispnB}. This inward running mode has a long amplification timescale
and a relatively low phase velocity. Therefore we do not expect a significant
contribution of this mode to the situation in the wind of model~A\@. For
model~B we find that Alfv\'en and fast magnetosonic waves have different phase
velocities now.  The fast magnetosonic waves are now amplified, too. But their
amplification timescale is much longer than for slow magnetosonic waves.
\section{Consequences from our model}
\begin{figure}
\setlength{\unitlength}{1 cm}
\begin{picture}(9,8.7)
\put(0,0){\epsfig{figure=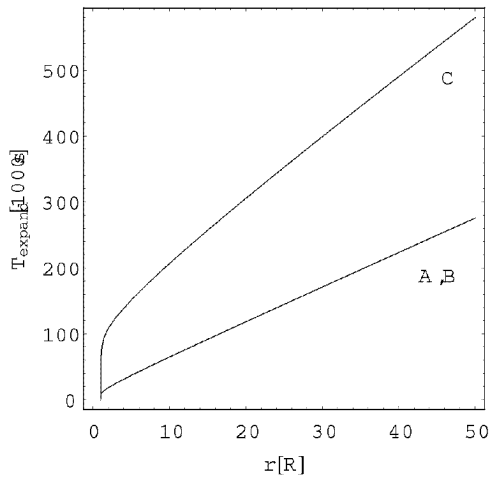}}
\end{picture}
\caption[]{\label{f:exp}Expansion time versus radius. 
Model~C has a slower 
acceleration close to the star and a lower terminal velocity. A fast
magnetosonic wave with an amplification timescale of about $2500\rm s$
(cf. Fig~\ref{f:modb:b}.a) grows by a factor of $\approx {\rm e}^{40}$, while
the wind expands from $r=1R$ to $r=1.5R$. The electron scattering opacity is
$^2/_3$ at $r=1.3R$. Therefore we can expect shocks already at this radius,
where line observation starts. The objections of Lucy (\cite{r:luc:84}) do not
change this situation qualitatively.}
\end{figure}
In this paper we do a linear stability analysis and show that the waves in the
stellar wind can help to understand the observations. In order to calculate
quantitative results, which can be compared with observations, it would be
necessary to analyze the detailed properties of the shocks resulting from the
waves found in this paper. But we can derive some estimates from our
calculations.

From Fig.~\ref{f:modb:b}.b we see that the phase velocity of the outward
running waves and possibly shocks in the rest frame of the star is at least
twice the terminal velocity of the unperturbed wind. In the outward running
waves the oscillations of $v_r$ and $\rho$ are in phase. Therefore a
significant amount of matter, but presumably not all matter will escape at this
or a higher velocity, if the outward running waves steepen into forward
shocks. The terminal velocity will then be overestimated at least by a factor
of two in the observation. For our model~C this would be $\vinfobs \approx
1500{\rm km\ s^{-1}}$. Krolik \& Raymond (\cite{r:kro:ray}) found that in a
nonmagnetic wind shock shells are running much faster than the phase velocity
of the waves. In an unperturbed magnetic wind model such a high value for
$\vinf$ combined with an reasonable high value for $\Mdot$ can only be obtained
with a very high magnetic field and fast rotation, which leads to a spin down
problem.

The influence of our model on $\Mdot$ is more difficult to estimate. We gain a
real factor on 28 in $\Mdot$ between our model A\&B and model~C even in the
unperturbed wind due to the driving force of the rotating magnetic
field. Furthermore the observation of $\Mdot$ is influenced by the clumping of
the wind matter. But to calculate the clumping factor
$<\!\rho^2\!>/<\!\rho\!>^2$ at large radii, where radio observations are made,
it would be necessary to do a nonlinear calculation including large radii,
because the waves steepen into shocks very rapidly due to the short
amplification timescales. This is beyond the linear model presented here.

Magnetic rotator models for hot stars are often criticized for their short spin
down times. We reduced this problem by using the equations of Biermann \&
Cassinelli (\cite{r:bie:csi}) for our unperturbed wind model~C\@. These
equations do not assume that the magnetic field is radial close to the
star. Therefore we find a smaller value for the Alfv\'enic radius $\RA$, which
controls as a lever arm the angular momentum loss. The magnetic field increases
the angular momentum loss in the equatorial plane compared to a nonmagnetic
star with the same rotation rate by $(\RA/R)^2$, which is 3.2 for our
model~C\@.

In order to decide, wheter the increased angular momentum loss would rule out a
rotating magnetic field as an important component of the wind of these stars,
one would need a detailed stellar evolution model. Such a model must e.g.\
consider the angular momentum structure and evolution inside the star, which is
yet unknown in the case of a strong magnetic field. It is also not clear,
wheter the magnetic field is present in the wind from the beginning of the
stellar evolution, or wheter it is generated later, as was suggested by
Biermann \& Cassinelli (\cite{r:bie:csi}) and Biermann (\cite{r:bie:97}) even
for stars with convective cores and radiative envelopes. The angular momentum
loss from the wind outside the equatorial plane is also not calculated yet. Due
to these serious uncertainties in this issue most people (including the
authors) use the very simple spin down formula of Friend \& MacGregor
(\cite{r:fri:mcg}, Eq.~38)
\begin{equation}\label{tau}
\tauJ = \frac{J}{\dot{J}} \approx \frac{3}{5}\frac{R^2}{\RA^2}\frac{M}{\Mdot}
=\frac{3}{5}\frac{R^2}{\RA^2}\tauM
\end{equation}
to roughly guess the spin down timescale of the star. This formula guesses the
angular momentum of the star by assuming, that the star is a {\it rigidly}
rotating {\it sphere} of {\it constant} density. All these three assumptions
are of course wrong. The angular momentum loss from outside the equatorial
plane is just guessed from the value in the equatorial plane using a factor of
$2/3$. Even if we could accurately calculate the current spin down $J/\dot{J}$
and mass loss $M/\Mdot$ timescales, $\tauJ<\tauM$ does not automatically mean
that post main sequence star does not rotate significantly. E.g. an artificial
nonmagnetic star $(R=\RA)$, wich fullfills the assumptions of Eq.~\ref{tau}
mentioned above , has no internal angular momentum transport and a constant
density over this lifetime, has $\tauJ = 3/5\tauM<\tauM$, but the rotation rate
$\Omega$ remains constant over the whole lifetime of such a star. The reason is
that the outer layers of the star have, due their larger average distance from
the rotation axis, more angular momentum per unit mass than inner layers in
spite of the rigid rotation. This example shows that Eq.~\ref{tau} is just
not accurate enough to verify or rule out a model with
$\tauJ\mbox{(Eq. \ref{tau})}\approx\tauM$ within an order of magnitude.  For
our model~C we find using Eq.~\ref{tau} $\tauJ=2.5\times 10^5\yr$ and
$\tauM=1.4\times 10^6\yr$.
\section{Conclusions}
In this paper we analyzed the interaction between a magnetic field and linear
waves induced by the radiative instability. We found both models complement
each other. The magnetic field suppresses the inward running waves, which
dominate in nonmagnetic winds. This may allow the outward running waves to
support the unperturbed wind as described by Koninx (\cite{r:kon}) and to form
high density shock shells running out at a high speed. These shock shells may
explain the high terminal velocities measured in winds of massive stars. The
outward running waves will also lead to an overestimation of $\Mdot$ due to
wind clumping and the overestimated $\vinf$.  Wind clumping also occurs without
a magnetic field. But in this case the resulting shock shells will run inward;
and the argument about the overestimated $\vinf$ would not apply. The
overestimated $\Mdot$ and $\vinf$ put unnecessarily strong restrictions on fast
magnetic rotator wind models. We argued that even for a magnetic field with
$B_{r0} = 500\rm G$ the spin down time is consistent with the lifetime of the
star inferred from the mass loss rate considering the uncertainties in the
stellar structure. From the observation of nonthermal radio emission in many OB
and Wolf-Rayet stars we know that these stars have a nonneglible magnetic
field. Further direct observations are necessary to infer the actual strength
of these fields.
 
We showed that a luminous fast magnetic rotator model plus
wind perturbations by waves or shocks can help to explain the observed high
values for $\Mdot$ and $\vinf$ without being ruled out by the spin down
problem. Further observation of the magnetic field and further theoretical work
on the evolution of stellar rotation are necessary to evaluate to role of
magnetic fields in winds of massive hot stars.
\begin{acknowledgements}
We thank Drs.\ J.\ P.\ Cassinelli, S.\ P.\ Owocki, and K.\ G.\ Gayley for their
useful comments which helped to improve this paper.
\end{acknowledgements}

\end{document}